# Ferroelectric $KNbO_3$ nanoplatelets for thermally driven pyrocatalytic hydrogen evolution and dye degradation

Salma TOUILI[a,b], Bouchra ASBANI[a], Youness HADOUCH[a,b,c], M'barek AMJOUD[b], Daoud MEZZANE[a,b], Nejc SUBAN[c,d], Hana URŠIČ[c,d], Nitul S. RAJPUT[e], Zdravko KUTNJAK[c,d], Brigita Rožič[c,d], Mustapha JOUIAD[a], Mimoun EL MARSSI[a,*]

a: LPMC, University of Picardie Jules Verne, Amiens 80039, France

b: IMED-Lab, Cadi Ayyad University, Marrakesh, 40000, Morocco

c: Jožef Stefan Institute, Jamova Cesta 39, Ljubljana, 1000, Slovenia

d: Jožef Stefan International Postgraduate School, Jamova cesta 39, Ljubljana, 1000, Slovenia

e: Technology Innovation Institute, Abu Dhabi, United Arab Emirates

* Corresponding author: mimoun.elmarssi@u-picardie.fr

**Abstract**

Day and night shift-induced thermal cycling offers a promising route toward free energy for green hydrogen production and dye degradation. Pyroelectric materials make this possible by converting temperature fluctuations into electrical charges that drive water splitting catalytic reactions and produce hydrogen fuel. Herein, we demonstrate an efficient pyrocatalytic hydrogen evolution reaction and Rhodamine B (RhB) degradation using ferroelectric potassium niobate ($KNbO_3$) perovskite nanoplatelets (KN-np) with an orthorhombic phase. Under thermal cycling between 20 and 50 °C, KN-np produced a high hydrogen yield of 680 $\mu mol\cdot g^{-1}$ over 30 thermal cycles, with an average hydrogen generation rate of approximately 22.67 $\mu mol\ g^{-1}$ per thermal cycle. Besides, KN-np pyrocatalytic activity enabled efficient degradation of the RhB dye up to 84 % after only 16 cycles with a kinetic rate constant of 0.11 per thermal cycle. Our findings show that the strong polarization and the excellent pyroelectric properties of KN-np are at the origin of the catalytic activity enhancement. This work lays the foundation for the future design of pyroelectric materials for clean energy production and environmental remediation.

## 1 Introduction

Energy, the lifeblood of all forms of life and movement, is unfortunately still predominantly derived from fossil fuels to power various sectors of global society and the economy. This reliance has exacerbated environmental crises, prompting a shift towards clean and renewable energy sources. Hydrogen fuel, produced through the eco-friendly water-splitting process, is a notable potential contributor to a sustainable energy economy [1]. A completely green energy cycle of supply, storage and use can be achieved by producing hydrogen from energy derived from renewable sources [2]. Renewable energy sources encompass a wide range of options that harness inexhaustible natural resources to produce electricity, including solar, thermal, wind, etc. [3].

Exploiting temperature fluctuations during the day-night cycle is a sustainable approach to renewable energy production [4]. Pyroelectric materials can effectively use such temperature variation. Only a spatial temperature oscillation can trigger pyroelectric materials to release electric charges and result in a flow of electrical energy [5]. Temperature oscillations-induced electrical charges can be efficiently exploited to drive the water-splitting reaction and produce green hydrogen, exploiting pyroelectric catalysis, a new path of combining electrochemistry and pyroelectricity [4].

In the pyroelectric catalysis process, temperature fluctuations induce the generation of oppositely charged polarities on the surfaces of the pyroelectric material [4]. This charge separation leads to forming a pyroelectric potential resulting from imbalanced bonds and their respective charges [6]. The resulting potential difference is a driving force for water oxidation and reduction reactions and for the decomposition of organic pollutant species [7]. Owing to their strong pyroelectric potential, induced by inherent spontaneous polarization, ferroelectric perovskite materials have attracted considerable interest in the field that leverages the pyroelectric effect [8]. Specifically, polycrystalline ferroelectric $ABO_3$-type oxide perovskite materials in bulk form have been extensively studied for pyroelectric-based applications [9–11]. However, the complete spontaneous polarization of bulk ferroelectric materials cannot be exploited effectively without applying poling treatment, as their randomly oriented domains limit the overall polarization efficiency [12]. In this respect, a significant improvement in the pyroelectric coefficient and spontaneous polarization values was achieved by subjecting the $Ba(Ni_{0.5}Nb_{0.5})O_3$ ceramic to different poling conditions [13]. This limitation had prompted the development of nano/microscale ferroelectric devices, providing new opportunities for high-performance, miniature applications due to their improved surface effects and controlled domains orientation allowing more efficient use of their pyroelectric properties. Building on this shift, several nanomaterials have been proposed for pyroelectric nanogenerators, including potassium niobate ($KNbO_3$) nanowires [14], lead zirconate titanate nanowires [15], polyvinylidene fluoride nanofiber membrane [16], and zinc oxide nanowires [6], among others. In this respect, chemically synthesized nanoparticles can grow along specific crystallographic axes, exhibiting spontaneous polarization and inherent crystalline anisotropy. Jung et al. [17] have reported that $KNbO_3$ nanorods can reach a piezoelectric coefficient of ($d_{33} \sim 55$ pm·$V^{-1}$). Conversely, the study by Li et al. [18] demonstrated that the pyroelectric effect of $K_xNa_{1-x}NbO_3$ nanocrystals considerably improved their photoelectrochemical properties for efficient water splitting (WS) reaction.

The development of nanoscale pyroelectric technologies has led to innovative applications, particularly in pyrocatalytic processes for wastewater remediation [7,19–22] and WS [23–25]. In this sense, pyroelectric nanomaterials can generate electrical charges useful for electrochemical redox reactions when repeatedly heated and cooled over a range of temperatures. For instance, polarized $BaTiO_3$ single crystals have been used for the direct WS reaction in direct contact with water under cold-hot thermal cycling between 40 and 70 °C [25]. Furthermore, the potential use of $Ba_{0.7}Sr_{0.3}TiO_3$ nanocubes for hydrogen evolution reaction (HER) has been explored by exploiting cold-hot thermal cycles between 25 and 50 °C [23]. In such cases, owing to their high specific surface area, finely dispersed water-suspended pyroelectric nano/microparticles are used to increase the charges available for hydrogen production. In addition to ferroelectric nanomaterials, two-dimensional pyroelectric black phosphorene has also been reported to undergo pyrocatalytic HER when thermally cycled between 15 and 65 °C [24]. Moreover, the feasibility of the pyrocatalytic WS process has also been theoretically proven by Kakekhani et al. [26]. Likewise, cadmium sulphide (CdS) has been used for the pyrocatalytic $H_2$ evolution by harnessing environmental cold-hot temperature fluctuations [27]. However, the abovementioned research and others [28–30] on pure pyroelectric catalysis for $H_2$ production remain limited and do not cover all the potential pyroelectric nano/micromaterials. This gap paves the way for exploring and introducing alkaline niobate-based nano/micromaterials as promising candidates to advance this field of study.

Among niobate-based materials, $KNbO_3$ stands out for this purpose since it presents a high pyroelectric coefficient of 93 μC·$m^{-2}$·$K^{-1}$ [31], excellent chemical stability, and high charge carrier mobility [32]. Its high pyroelectric coefficient demonstrates its ability to leverage thermal changes as an energy source and highlights its potential as a powerful catalyst for efficient, temperature-driven WS reactions. Furthermore, its suitable band position for oxidation and reduction reactions in WS and water remediation made it among the best candidate for photocatalysis, piezocatalysis, and their synergistic effects [33–36]. This unique band alignment enhances its applications in photocatalysis and piezocatalysis and makes it particularly desirable for pyrocatalytic WS and dyes decomposition. It has been reported that $KNbO_3$'s good crystallinity and rich surface active sites enhance charge transfer and surface redox reactions, respectively [37]. Besides its non-toxicity, eco-friendliness, and ease of synthesis, it also exhibits high dispersibility in water required for such catalytic applications. Therefore, whether on the grounds of environmental protection or application prospects, it is of great interest to investigate, $KNbO_3$ material for the pyroelectric catalysis process for the first time in our knowledge.

In this study, we investigated $KNbO_3$ nanoplatelets for the pyrocatalytic HER and degradation of RhB under the stimulation of cold-hot cycles between 20 to 50 °C. $KNbO_3$ was synthesized using a simple hydrothermal method, followed by a systematic set of characterizations. Furthermore, we comprehensively analyzed its pyro-electrochemical properties, providing valuable insights into its performance.

## 2 Experimental

### 2.1 Synthesis of KN nanoplatelets

A simple hydrothermal method was used to synthesize pure $KNbO_3$ nanoplatelets (KN-np). Initially, 1 g of niobium oxide was slowly added to a highly concentrated aqueous potassium hydroxide solution (16 M). The resulting mixture was then stirred vigorously at room temperature for 1 hour, transferred to a Teflon-lined autoclave (100 ml), and reacted at 180 °C for 24 hours, as represented in Fig. 1.

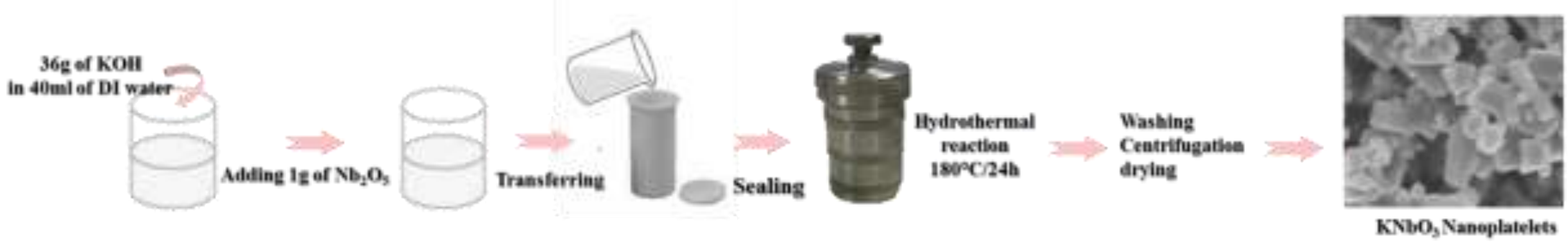

Fig. 1. Synthesis process of KN-np.

### 2.2 Structural, microstructural, and piezo-response characterizations

The crystalline structure of KN-np was characterized at room temperature by Panalytical X-Pert Pro X-rays diffractometer, Cu Kα radiation (λ = 1.54 Å) ranging from 20° to 80° with angle steps of 0.02° and a scan speed of 2° per minute. Raman spectroscopy of powder specimen was carried out at room temperature in the wavenumber range of 50 - 1000 $cm^{-1}$ under a green excitation laser of 532 nm using a Renishaw in Via Reflex Raman spectrometer equipped with an Edge-filter. The microstructure and morphology of the powder sample were analyzed by a field emission scanning electron microscope (FE-SEM, JSM-7600F, JEOL, Japan), transmission electron microscope (TEM) and high-resolution field emission transmission electron microscope (HRTEM), using a Titan G2 (ThermoFisher Scientific) Cs-corrected image TEM system, operating at 300 kV. The piezoelectric performance was analyzed using piezo-response force microscopy (PFM). Before the PFM measurements, the KN powder was embedded in polymer resin (Versocit, Struers, Ballerup, Denmark) to prevent the particles from sticking to the PFM tip. After curing for 30 minutes and polishing using standard metallographic technique, the sample was examined with an atomic force microscope (AFM, Asylum Research, MFP-3D, Santa Barbara, USA) equipped with a piezo-response force module (PFM). A conductive titanium/platinum-coated silicon tip with an aluminium-coated reflective side and a radius of curvature of ~15 nm (OMCL-AC240TM-R3, Olympus, Japan) was used for scanning. Due to the embedding of the powder in the epoxy resin, the electric field was applied to the sample via the conductive PFM tip in the virtual ground state, similar to that previously reported [38,39]. The out-of-plane piezo-response amplitude images were recorded in dual AC-resonance tracking (DART) mode at an AC amplitude signal of 5 V and ~350 kHz. The PFM amplitude and phase hysteresis loops were recorded in the switching spectroscopy mode with a pulsed DC step signal and a superimposed drive AC signal [40]. The waveform parameters were as follows: the sequence of increasing DC electric field steps was driven at 20 Hz with a maximum amplitude of 30 V; the frequency of a triangular envelope was 200 mHz; an overlapping sinusoidal AC signal with an amplitude of 2 V and a frequency of ~350 kHz was used. Three cycles were recorded in an off-electric field mode.

### 2.3 Pyro-electrochemical and optical measurements

All the pyro-electrochemical measurements were conducted on an electrochemical workstation (VSP-3e) with a standard three-electrode cell containing a working electrode, a Pt wire as a counter-electrode, and an Ag/AgCl as a reference electrode. The cell has a flat quartz window, allowing infrared heating of the front of the working electrode. An aqueous solution of $Na_2SO_4$ (0.1 M, pH = 7) was used as the electrolyte. The spray coating technique was used to coat the working electrode with KN-np using an indium tin oxide (ITO) glass substrate [41]. This was achieved by ultrasonically dispersing the KN-np in ethanol solvent with a mass concentration of 0.1 $g \cdot L^{-1}$, then depositing it onto the ITO substrate at 350 °C at a rate of 1 $mL \cdot min^{-1}$. The surface area of the electrode covered with the sample is 1 $cm^2$. The pyroelectric current generated by the sample was recorded under the stimulation of a range of temperature gradients. The electrochemical impedance spectroscopy (EIS) was conducted over a frequency range of 10 mHz to 1 MHz. The Mott-Schottky plot was determined at a frequency of 1 kHz. The optical properties of the hydrothermally synthesized KN-np were evaluated by UV-vis spectroscopy using a Shimadzu UV-2600 spectrophotometer for wavelengths ranging from 300 to 800 nm.

### 2.4 Pyro-electrolytic HER measurement

In a simple experiment, 20 mg of KN-np was dispersed in 20 ml of a mixture of deionized water/methanol (20 vol% methanol). The resulting aqueous suspension was sealed in a 35 ml borosilicate tube, which was then evacuated and purged with nitrogen for 5 minutes to remove all the residual air. Thermal cycles were carried out in a range of between 20 and 50 °C using water bath devices. Accordingly, the borosilicate tube was transferred between heating and cooling water baths. The entire catalytic reaction was conducted under dark conditions to prevent the light contribution. Hydrogen gas detection was performed by carefully and intermittently inserting a Unisense hydrogen microprobe sensor into the borosilicate tube at 20 °C thermal cycle to prevent any disturbance of temperature measurements.

### 2.5 Dye decomposition experiments

RhB degradation experiment was conducted in a glass beaker after 1 hour of stirring to achieve adsorption-desorption equilibrium between the dye and the KN-np catalyst. Subsequently, the beaker containing 20 mg of KN-np dispersed in 20 ml of RhB (5 $mg \cdot L^{-1}$) was placed in the center of a water bath with very low stirring speed to undergo the thermal cycling process in the dark. 2 ml of RhB dye was taken at each thermal cycle and centrifuged to separate it from the catalyst. The concentration of the RhB dye solution was determined by measuring the absorption peak at 554 nm. To investigate the pyrocatalytic mechanism underlying RhB degradation, isopropanol (IPA) was used as a hydroxyl radical ($^{\bullet}OH$) scavenger [42], methanol as a hole ($h^+$) scavenger [43], and ascorbic acid (AA) as a superoxide radical anion ($^{\bullet}O_2^-$) quencher [44]. The appropriate amount of scavenger was added to the dye-catalyst mixture before initiating the pyrocatalytic dye decomposition experiment.

# 3 Results and discussion

## 3.1 Structural and microstructural properties

The crystal structure of KN-np was studied at room temperature by XRD and Raman analyses. Fig. 2(a) shows the XRD patterns of the as-prepared KN-np without any additional calcination treatment that causes potassium oxide ($K_2O$) to evaporate at temperatures above 800 °C, which results in the formation of undesirable phases when the K/Nb ratio differs from unity [31]. According to the JCPDS database (card no.:32–0822), all diffraction peaks are indexed to the orthorhombic phase of KN-np (space group Amm2) with lattice parameters of a = 3.9829 Å, b = 5.6624 Å, c = 5.7270 Å, and a cell volume of 129.1605 $Å^3$ [45]. The apparent splitting of the (022) and (200) crystal planes, shown in the enlarged view at 2θ = 45° in Fig. 2(b), confirms the crystallization in an orthorhombic structure [46]. The sharp diffraction peaks reflects the high crystallinity of the hydrothermally synthesized KN-np [47]. Compared with XRD, Raman spectroscopy is highly sensitive to lattice distortions within a specific local structure [45]. The Raman spectrum of the as-prepared KN-np is shown in Fig. 2(c) featuring typical Raman peaks around 190, 257, 280, 296, 531, 591, and 831$cm^{-1}$. It has been reported that the sharp low-frequency peak located at 190 $cm^{-1}$ corresponds to the ($B_1$, $B_2$)TO mode [48]. The three overlapping bands appearing at 257, 280, and 296 $cm^{-1}$ are assigned to modes $B_1$(TO), $A_1$(TO), and $A_1$(TO, LO), respectively [49]. The bands positioned at 531 and 591 $cm^{-1}$ belong to the $B_1$(TO) and $A_1$(TO) modes, respectively [50]. The last band observed at 831 $cm^{-1}$ could be assigned to the $A_1$(LO) mode [49]. All these phonon modes confirm the orthorhombic structure of KN-np and agree with previous reports in the literature [51]. The band peak observed at 190 $cm^{-1}$ is associated with a long-range polar order in KN-np [52]. The above three overlapping bands can be attributed to the O-Nb-O symmetrical bending vibration of the $NbO_6$ octahedron [53]. The high-frequency bands observed at around 531 and 591 $cm^{-1}$ are attributed to the symmetrical Nb-O stretching modes of the $NbO_6$ octahedron [54]. In addition, the weak band at 831 $cm^{-1}$ may correspond to the combinatorial band of the Nb-O stretching mode of the tetrahedral $NbO_4$ unit induced by strong distortion of the $NbO_6$ octahedra [54].

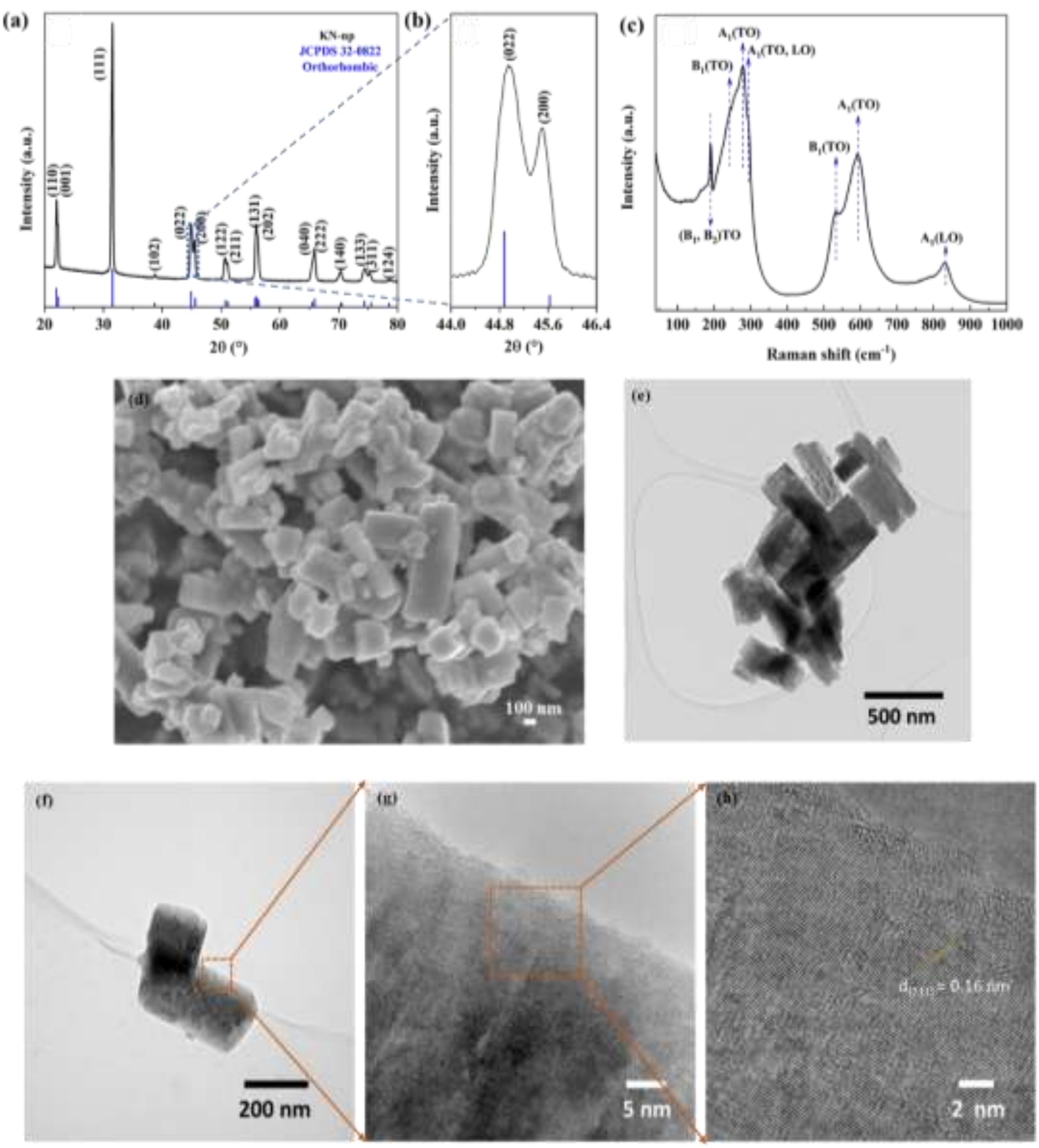

Fig. 2. (a) XRD diagram of KN-np; (b) enlarged view of XRD peak centered around 45.5°; (c) Raman spectrum of KN-np captured at room temperature; (d) SEM photograph; (e-h) HRTEM images of KN-np.

SEM and TEM experiments were used to examine the morphology of the resultant KN-np. SEM and low-magnification TEM images of the KN-np shown in Fig. 2(d), (e) reveal their nanoplatelet morphology. The tendency to aggregation observed in the synthesized KN-np specimen is due to the known high surface energy of the nanoparticles. The crystal structure of the sample was examined in greater detail by HRTEM and enlarged in Fig. 2(h). The observed lattice spacing of 0.16 nm, correspondes precisely to the (211) crystallographic plane, in agreement with the crystal parameters of orthorhombic KN-np. Overall, these findings are consistent with the XRD and Raman results, confirming the high purity and crystallinity of the synthesized KN-np.

### 3.2 Ferroelectric and piezoelectric properties

Piezo-response force microscopy (PFM) was used to explore the local piezoelectric properties of KN nanopowders. The AFM topography height and deflection images and PFM out-of-plane amplitude and phase images of KN-np embedded in the epoxy matrix are shown in Fig. 3. The bright contrast of the KN-np particles in the PFM amplitude images indicates their piezoelectric activity compared to the dark contrast of the non-piezoelectric epoxy. Even the ferroelectric domain structure is observed in larger KN-np particles (Fig. 3(c)). The local PFM amplitude and phase hysteresis loops were also measured using PFM switching spectroscopy mode. The typical ferroelectric/piezoelectric behavior indicates effective polarization switching in KN-np, resulting in typical shapes of hysteresis loops for ferroelectric material (Figs. 3 (e), (f)). Good pyroelectric properties could, therefore, be expected for this powder.

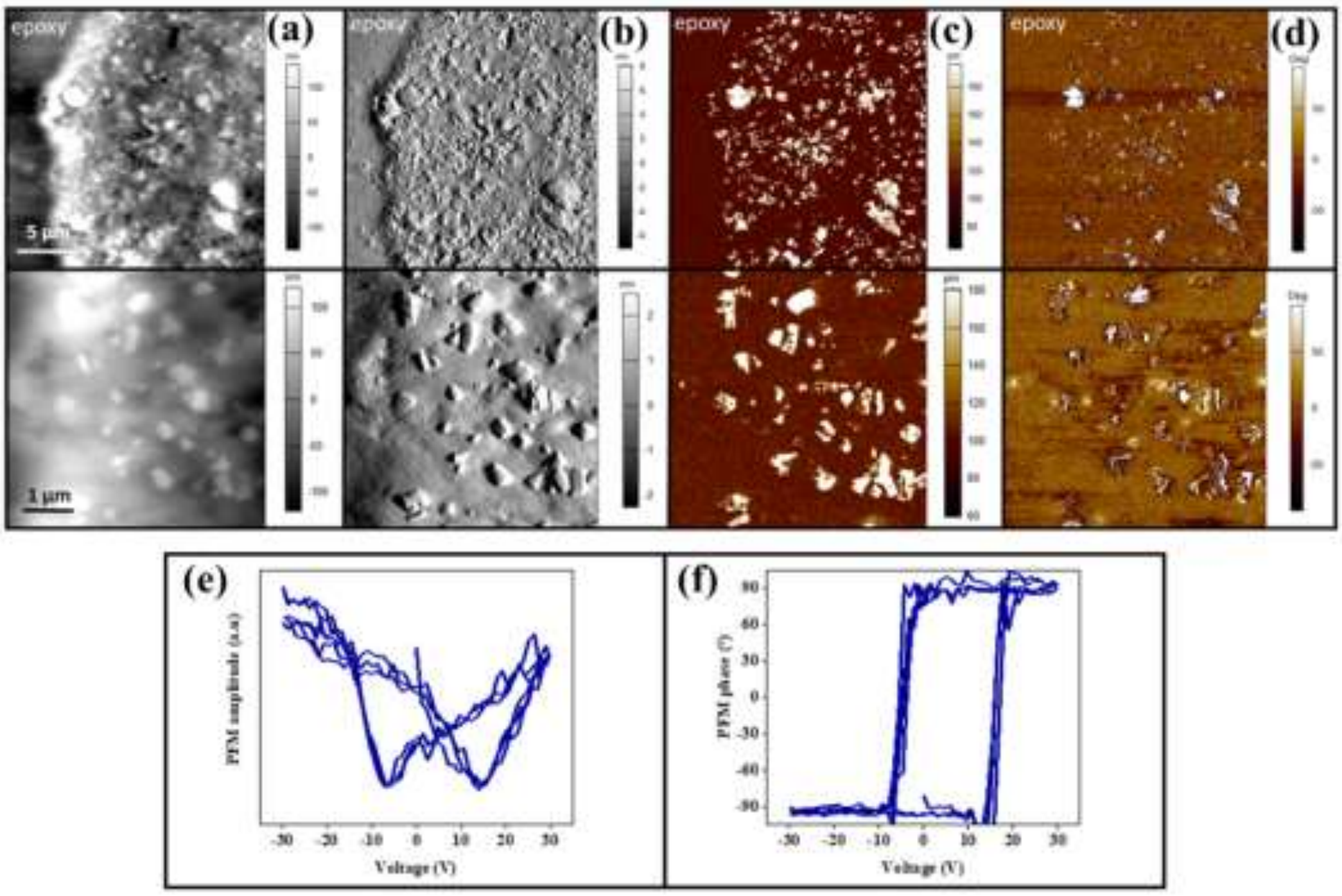

Fig. 3. AFM topography (a) height and (b) deflection images; PFM out-of-plane (c) amplitude and (d) phase images; PFM (e) amplitude and (f) phase hysteresis loops of KN-np embedded in the epoxy matrix.

### 3.3 Pyro-electrochemical properties

Pyroelectric current measurements are crucial for determining pyroelectric parameters for bulk materials or nanoparticles. The zero-bias pyroelectric current generated by KN-np was measured under cyclic cold-hot fluctuations at two thermal frequencies and represented in Fig. 4(a), (b). The thermal frequencies of 0.1 and 0.2 Hz correspond to temperature variations ($\Delta T$) 0.35 and 0.16 K, respectively. For all thermal frequencies, the pyro-current exhibits a sharp rising edge, a stable plateau, and a falling edge; this behavior is identical to that of previously reported pyroelectric materials [24,55]. Specifically, a rapid and pronounced transient pyroelectric current appeared with a positive current value when $\Delta T > 0$, whereas a prominent current with a negative value was detected when $\Delta T < 0$ (Fig. 4(c)). The measured pyroelectric current peaks were 351 $nA\cdot cm^{-2}$ and 400 $nA\cdot cm^{-2}$ under $\Delta T$ of 0.16 and 0.35 K, respectively. The observed rise in the pyro-current peak as the frequency of the thermal cycling decreases can be explained by the increase in the pyroelectric charges released ($Q_{pyro}$), as clearly shown in Fig. 4(a), (b). Larger heat transfer and homogeneous temperature distribution could be expected at a lower frequency (0.1 Hz) due to exposing the pyroelectric nanoparticles to the heat source for approximately longer periods [28]. The induced pyroelectric charges $Q_{pyro}$ can be expressed as:

$$Q_{pyro} = p.A.\Delta T \quad (1)$$

Where p and A are the pyroelectric coefficient and the coated surface of the KN-np electrode, respectively. According to this equation, $Q_{pyro}$ generated is directly proportional to the temperature change $\Delta T$, with a larger $\Delta T$ being critical for larger charge release, which is consistent with the experimental results. The frequency of thermal cycling causes a change in the polarization of the pyroelectric material due to the change in $\Delta T$, given that $dP = p\cdot\Delta T$. This change in polarization is essentially due to separate polarization charges being released with opposite polarities on the KN-np polar surfaces under temperature alternation. This creates potential differences and band bending caused by the disequilibrium of polarization and compensation charges [7]. In this way, temperature-induced polarization can create a static pyroelectric potential difference and a time-dependent pyroelectric current. The pyroelectric coefficient can be expressed as follows:

$$p = \frac{I}{A\cdot dT/dt} \quad (2)$$

Where I and dT/dt represent the pyroelectric current and the thermal cycling rate, respectively. The calculated pyroelectric coefficient of approximately 13 $\mu C\cdot cm^{2}\cdot K^{-1}$ is an order of magnitude higher than that previously reported for bulk KN [31]. During the slight temperature variations studied, the pyroelectric coefficient p was basically a constant. Typically, the higher the thermal cycle rate dT/dt, the greater the pyroelectric current generated at the material's surface. However, in this pyroelectric current measurement, a relatively slow dT/dt of around 0.03 $°C\cdot s^{-1}$ was adopted to prevent interference from the thermoelectric effect [27]. Despite this, the KN-np pyroelectric catalyst achieved a high pyroelectric current of 400 $nA\cdot cm^{-2}$, and a large pyroelectric coefficient of 13 $\mu C\cdot cm^{2}\cdot K^{-1}$. At the same thermal rate of 0.03 $°C\cdot s^{-1}$, a pyroelectric coefficient of 0.47 $\mu C\cdot cm^{2}\cdot K^{-1}$ was obtained by CdS nanorods used for the pyrocatalytic HER [27]. The strong pyroelectric current signal obtained by the KN-np catalyst could be related to the formation of a built-in electric field that arises from the high polarization of orthorhombic KN-np during the thermal cycling process, which facilitates the charge separation and tranfer process [56].

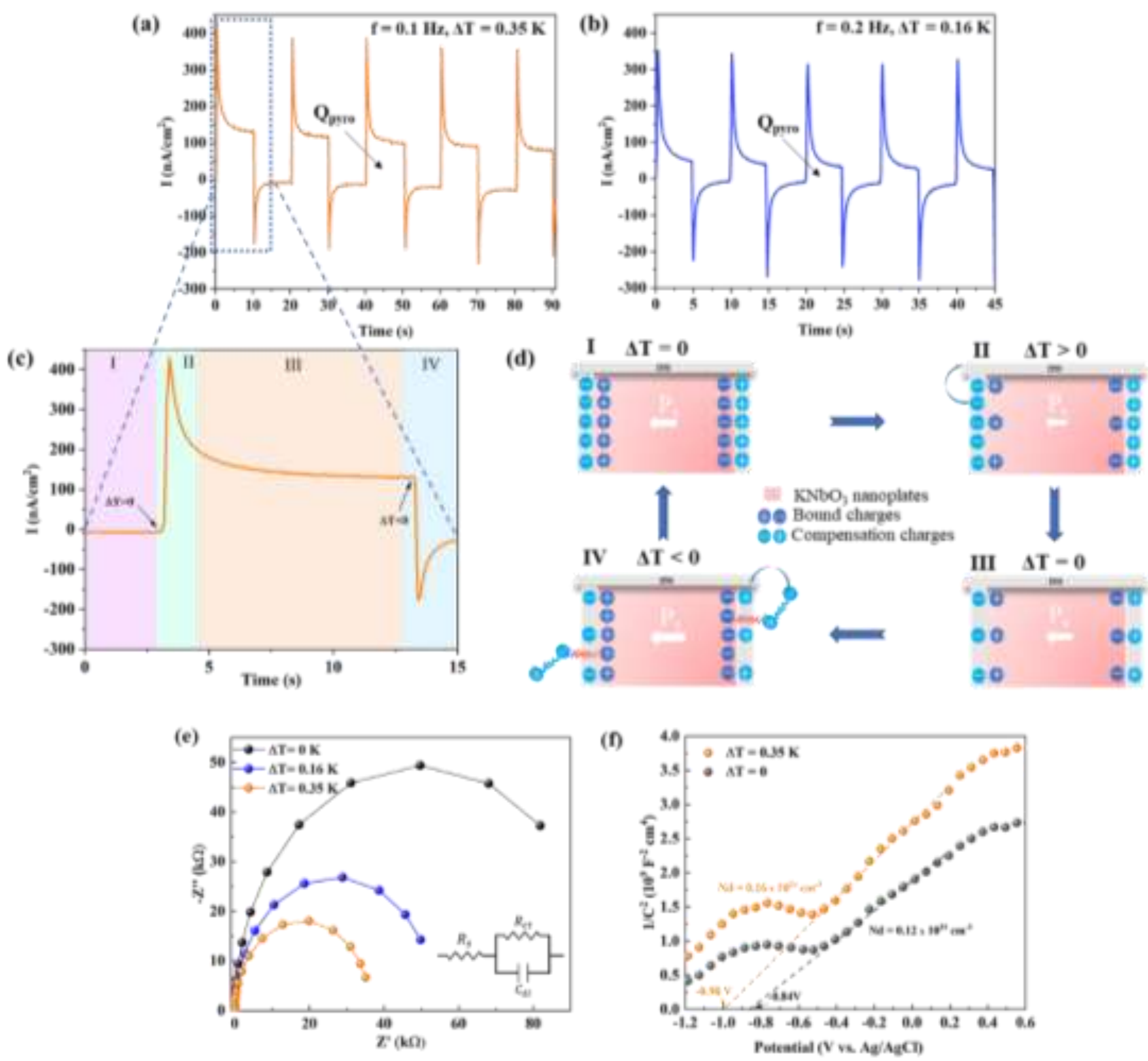

Fig. 4. Pyroelectric current measurements at (a) 0.1 Hz and (b) 0.2 Hz thermal frequencies; (c) a zoomed view of a thermal cycle from the figure (a); and (d) the proposed pyroelectric current generation mechanism; (e) EIS plots of KN-np; (f) Mott–Schottky plot of KN at 1 kHz.

The study of pyroelectric current is essential for accurately determining the pyroelectric coefficient and gaining insight into the mechanisms underlying current generation. To investigate the physical mechanism of the pyroelectric effect, one cycle of the I-t curve was divided into four stages, labelled I, II, III, and IV, as shown in Fig. 4(d). At constant temperature, the KN electrode maintains a thermodynamic equilibrium with the electrolyte, where the compensating charges balance the bound charges generated by spontaneous polarization at both ends of the electrode ($\Delta T = 0$, phase I). Given the primary pyroelectric effect, temperature fluctuations lead to a redistribution of charge carriers at the electrode surface [57]. During thermal excitation ($\Delta T > 0$, phase II), the intensity of spontaneous polarization decreases due to the disruption of the random oscillation state of the electrical dipole moments, resulting in a decrease in bound charges and a release of some compensation charges [58]. Accordingly, the free compensation charge carriers flow towards the counter-electrode, generating a positive pyroelectric current. When the temperature stabilizes ($\Delta T = 0$, phase III) and a suitable concentration of compensating charge carriers accumulates at the surface, the KN-np polarization field is neutralized, establishing a new equilibrium and reducing current density. In this new state of equilibrium, the higher current density observed compared to phase I is due to the additional piezoelectric polarization resulting from the heat-deformation conversion, known as the secondary pyroelectric effect. Finally, when $\Delta T < 0$ in phase IV, the intensity of KN-np polarization is magnified due to the creation of more bound charges, accompanied by the generation of additional compensation charges on the surfaces and a reverse pyroelectric current [59]. Overall, current generation without applying bias voltage demonstrates that the KN-np has an inherent pyroelectric current. Stated differently, a temperature shift can spontaneously polarize orthorhombic KN-np, thereby activating their pyroelectric effect.

It is possible to ascertain the charge generation, separation, and transfer process by investigating the KN-np sample's electrochemical characteristics under various conditions. EIS is performed to further examine the carrier transfer behavior of KN-np pyro-catalyst. Fig. 4(e) shows the fitted EIS Nyquist plots for different temperature fluctuations $\Delta T$. The inset of this figure shows the equivalent circuit, where $R_s$, $R_{ct}$, and $C_{dl}$ correspond respectively to the resistance of the solution, the charge transfer resistance at the semiconductor-electrolyte interface, and the capacitance of the electrochemical double layer. The applied thermal field effectively reduces the solution's resistance while minimizing its semicircle. This decrease in solution resistance demonstrated by the Nyquist plots confirms that the temperature field induced a high concentration of compensating charge carriers at the surface of the KN-np. In this way, the built-in electric field induced by the pyroelectric effect stimulates an increase in the concentration of carriers and contributes directly to the separation and transfer of charges on the surface of the KN-np sample [60].

The bands' alignment at the semiconductor/electrolyte interface is a major determinant of charge transfer [61]. Mott-Schottky plot at 1 kHz is used to determine the flat-band potential ($V_{fb}$) during the interfacial carrier migration. The n-type behavior of the KN-np is illustrated by its positive slope, as illustrated in Fig. 4(f). By applying an electrical potential, the flow of numerous additional electrons will decrease the band-bending and move the Fermi energy level slightly upwards [62]. The negative shift in the $V_{fb}$ with temperature stimulation is attributed to increased carrier availability, improved charge separation, and increased donor density resulting from the built-in electric field generated in the pyroelectric material [60]. Moreover, this shift indicates enhanced band bending at the semiconductor/electrolyte interface [63].

The intercept on the horizontal axis of the Mott-Schottky plot at non-cycled conditions indicates a $V_{fb}$ of -0.84 V vs. Ag/AgCl, which can be converted into the reversible hydrogen electrode (RHE) scale using the Nernst equation (3), resulting in -0.23 V vs. RHE.

$$E_{RHE} = E_{Ag/AgCl} + 0.059 \times pH + E^{\circ}_{Ag/AgCl} \quad (3)$$

With $E°_{Ag/AgCl}$ = 0.1976 V vs. RHE at 25ºC, pH = 7. For undoped n-type semiconductors, the $V_{fb}$ is roughly 0.3 V below the conduction band (CB) minimum [64]. Thus, the minimum of the CB of KN is -0.53 V vs. RHE at pH 7. Overall, these findings demonstrate that the KN-np catalyst can initiate a redox reaction triggered by temperature changes, making it a promising candidate for applications involving HER and wastewater treatment.

Nanostructured piezoelectric materials are usually considered to have more significant piezoelectric potential than their bulk counterparts owing to their ability to deform readily [35]. Similarly, incorporating of pyroelectric nanostructures into devices has led to a major change in the energy harvesting industry [65]. Owing to its suitable polarization field, KN nanoparticles have been widely utilized in photoelectrochemical cells for piezo-photocatalytic WS and wastewater remediation processes. Hydrothermally synthesized KN nanosheets demonstrated improved piezo-photocatalytic degradation efficiency of organic dyes compared to their nanocube counterparts, due to their greater piezo-ferroelectric response as demonstrated by PFM analysis [35]. Additionally, it was revealed that by adjusting the phase transition temperature through the size engineering process, a high pyroelectric response was achieved in ferroelectric nanowires [66]. KN nanowires embedded in polydimethylsiloxane polymer was successfully used as pyroelectric nanogenerators for direct thermal energy harvesting and sunlight-induced heat harnessing [14]. However, to our knowledge, there is not yet any reported literature on the pyrocatalytic properties of nanostructured $KNbO_3$ for HER and RhB dye decomposition.

### 3.4 Pyrocatalytic HER and mechanism

Leveraging natural temperature fluctuations, such as the day/night cycle, for hydrogen production is highly desirable. For this reason, the thermal cycling process in this study is designed to operate in the temperature range of 20–50 °C, optimizing the system for practical and sustainable energy production. In addition, this temperature range is a balanced compromise for achieving enough ΔT and a manageable cycle time for efficient HER. Several studies have already adopted this range of temperature variation for pyrocatalytic applications using alkaline niobate-based materials [60,67]. The ideal curve of temperature fluctuation is shown in Fig. 5(a), with the flow of the reaction being temperature increase, heat conservation, temperature decrease and heat conservation, with each step lasting 150 s. To illustrate the pyrocatalytic effectiveness of the KN-np catalyst, the HER during the pyrocatalytic process of WS under variations in environmental temperature change was measured in the presence of methanol used as a sacrificial agent. As shown in Fig. 5(b), the hydrogen yield increases with the thermal cycling process, reaching 680 $\mu mol \cdot g^{-1}$ after 30 cycles. Under these conditions, the hydrogen production rate of KN-np is 22.67 $\mu mol \cdot g^{-1}$ per thermal cycle. As a classical ferroelectric material, KN-np has a high remanent polarization in its bulk form, approximately 18 $\mu C \cdot cm^{-2}$, as demonstrated elsewhere [68,69]. Furthermore, the PFM phase hysteresis loop demonstrated the strong nanoscale remanent polarization of the KN-np used for HER (Fig. 3(f)). This high polarization is the main reason for this material's high pyrocatalytic hydrogen production. Essentially, the internal electric field associated with remanent polarization promotes the formation and separation of pyro-generated electron-hole pairs when the material undergoes temperature fluctuations. Otherwise, it has been reported that the rapid recombination of positive and negative charges in the pyrocatalytic process reduces the hydrogen production yield, making adding a sacrificial electron donor essential [23,24,29,30]. In this respect, the addition of methanol as a sacrificial agent to the $Ba_{0.7}Sr_{0.3}TiO_3$ pyro-catalyst solution led to a significant increase in hydrogen production yield, rising from 0.75 μmol·g$^{-1}$ to 46.89 mmol·g$^{-1}$ after 36 thermal cycles [23]. This highlights the essential role of methanol in improving charge separation and enhancing the overall efficiency of the pyrocatalytic process [23].

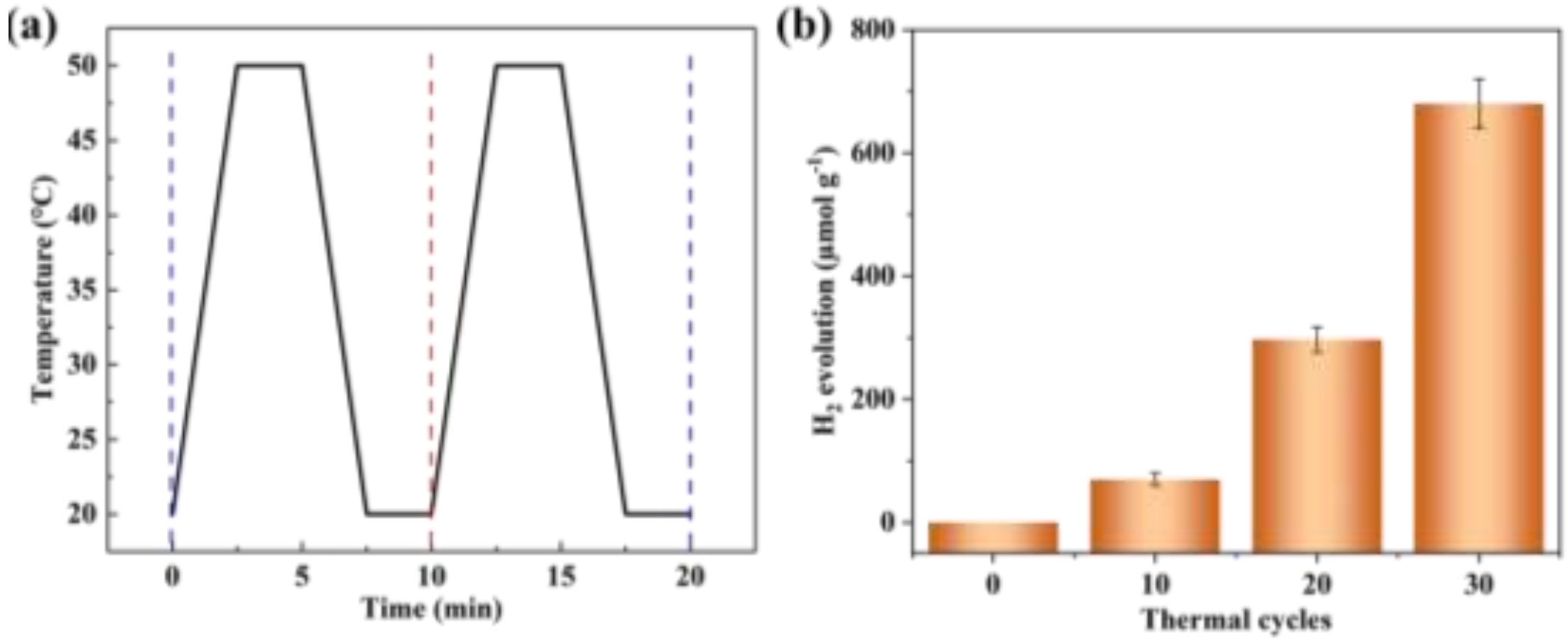

Fig. 5. (a) The ideal temperature curve of the pyrocatalytic evolution of hydrogen; (b) The pyrocatalytic HER of KN catalyst under temperature fluctuations.

Table 1 summarizes the current investigations into the pyrocatalytic HER using different classes of materials in different environmental temperature fluctuations. The yield of hydrogen production varies according to the materials used and the conditions, range and cycles of temperature fluctuation. Compared with hydrogen production with pyroelectric nanopowders, using an external pyroelectric device as an energy supplier for HER has been demonstrated by Zhang et al. [28] using lead zirconate titanate sheet. The high hydrogen yield generated by KN-np underscores their potential as a sustainable external energy source for future hydrogen production. Overall, this efficient technique of producing hydrogen, driven by hot-cold temperature variations, could significantly advance the field of renewable energies, particularly through exploring diverse pyroelectric materials.

Table 1. Comparison of the pyrocatalytic performance of different pyroelectric catalysts for HER.

| Pyro-catalyst | Cycles | Cycling temperature (°C) | Sacrificial agent | $H_2$ yield (μmol·g$^{-1}$) | Refs. |
|---|---|---|---|---|---|
| $KNbO_3$-np | 30 | 20–50 | Methanol | 680 | This work |
| $Ba_{0.7}Sr_{0.3}TiO_3$ | 36 | 25–50 | Methanol | 46.89 | [23] |
| B-phosphorene | 24 | 15–65 | Methanol | 540 | [24] |
| CdS | 36 | 25–55 | Lactic acid | 154.8 | [27] |
| SiC | 20 | 27–60 | Methanol | 32.84 | [29] |
| $Si_3N_4$ | 20 | 27–60 | Methanol | 12.32 | [30] |

In order to analyze the mechanism of HER, the KN band structures were considered. The optical spectrum and Tauc plot of KN-np are depicted in Fig. 6(a) and (b), respectively. Fig. 6(a) shows the UV light absorbance edge, demonstrating its absorption is below 380 nm. The Tauc equation ($(\alpha h\nu)^2 = (h\nu - Eg)$) was used to determine the direct band gap of KN-np [70]. Where α represents the absorption coefficient, hν refers to the energy of the incident photon, $E_g$ represents the band gap of the sample, and n = 2 for the direct transition [70]. The $E_g$ value can be estimated from the intersection of the tangent of $(\alpha h\nu)^2$ versus photon energy (hν) to the x-axis, as shown in Fig. 6(b). The calculated $E_g$ value is 3 eV, lower than that reported in a previous study (3.2 eV) for KN orthorhombic nanowires [71].

The energy band alignment of KN is illustrated in Fig. 6(c). At pH 7, the conduction band level for the $H^+/H_2$ redox couple is -0.41 V vs. RHE, which is higher than the conduction band minimum of KN (-0.53 V vs. RHE), enabling the reduction of water to produce hydrogen. Thus, the catalyst's good band alignment with water's potential redox indicates its thermodynamic ability for the WS reaction. Moreover, dynamic control of carrier energy in ferroelectric materials that include KN-np derives from their inherent spontaneous polarization, which induces internal electric field that directly influences the near-surface band structure [63]. Under the effect of temperature fluctuations, this polarization changes through the pyroelectric effect, leading to a modulation of the built-in electric field and, consequently, modifying the degree of band bending [7].

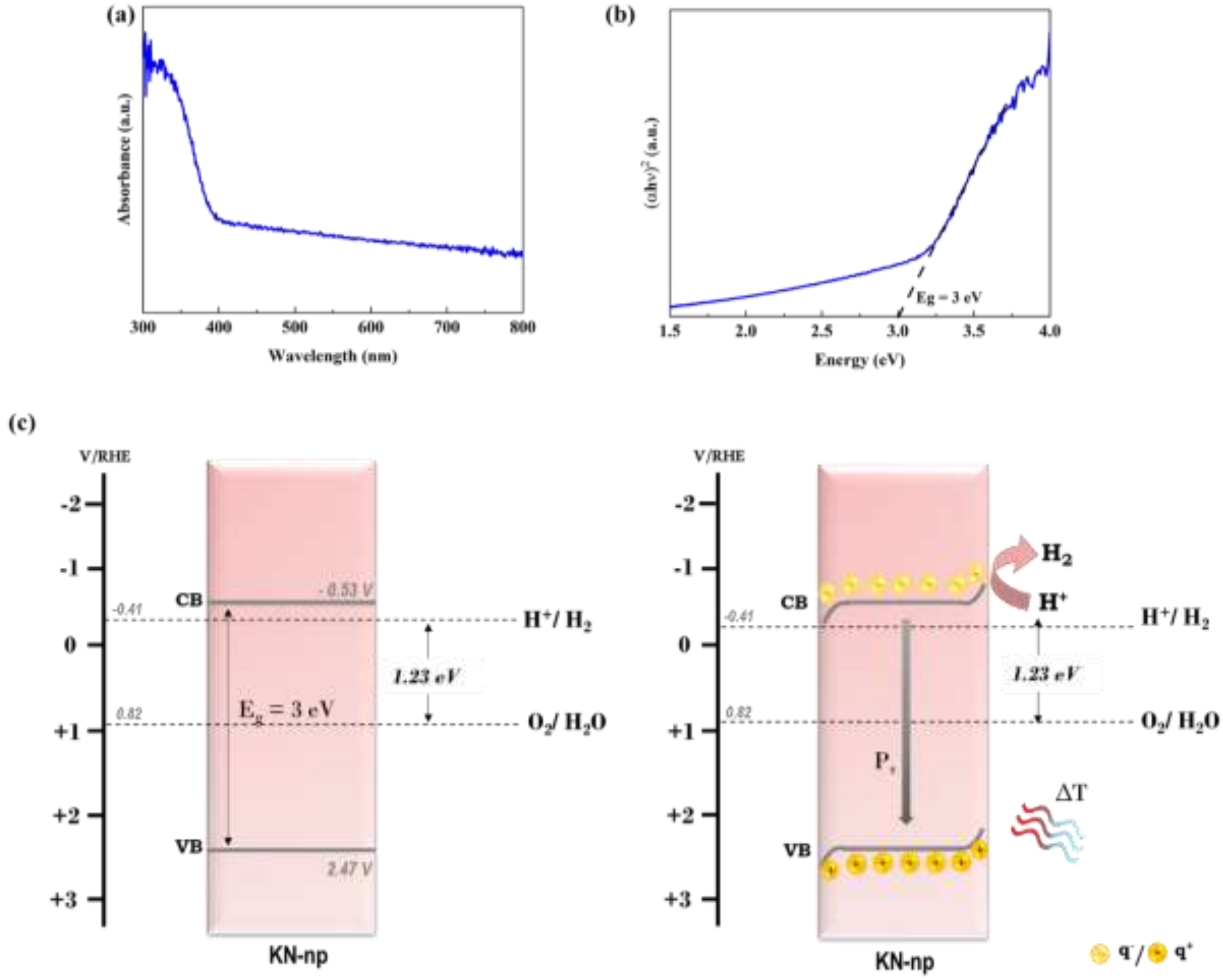

Fig. 6. (a) UV-vis spectrum of KN sample; (b) corresponding plot $(\alpha h\nu)^2$ vs. energy h$\nu$; (c) Schematic of the energy band level of KN-np; (d) The energy bands switch under the effect of the strong pyroelectric field induced by the thermal cycling.

The mechanism of pyrocatalytic HER using KN-np is depicted in Fig. 6(d). In pyroelectric KN-np, a spontaneous polarization field is generated in response to temperature fluctuations (ΔT). This polarization field creates an internal electric field that facilitates the separation of thermally generated charge carriers [4]. As a result, positive and negative charges are separated more efficiently, suppressing recombination and improving interfacial charge transfer, as described by equation (4) and detailed in the pyro-electrochemical section. The pyro-induced positive charges $q^+$ can facilitate the oxidation of water molecules adsorbed on the KN-np surface, leading to the generation of hydrogen ions $H^+$ and oxygen $O_2$ [72]. Subsequently, the hydrogen ions $H^+$ interact with the pyro-induced negative charges $q^-$ on the surface, producing hydrogen gas $H_2$, as outlined in equation (5). Otherwise, the positive charges $q^+$ generated have been suggested to react with the sacrificial agent methanol, generating $H^+$ and an intermediate hydroxyalkyl radical ($^{\bullet}CH_2OH$) [73]. The downward bending of the band observed at the interface due to the polarization field moves the edge of the conduction band towards a more negative potential (Fig.6 (d)). This is thermodynamically favorable for negative charge transfer to the $H^+$, facilitating the evolution of hydrogen in pyrocatalysis. The negative shift in $V_{fb}$ thus confirms the role of the pyroelectric effect in modifying band alignment and promoting charge separation [63]. Furthermore, cyclic charge generation during temperature fluctuations enabled continuous HER, making KN an effective pyro-catalyst. Overall, the dynamic coupling between spontaneous polarization, the generation of a built-in electric field, and the enhancement of the band bending creates an interfacial environment favorable to the pyrocatalytic HER.

$$KNbO_3 \xrightarrow{\Delta T} KNbO_3 + q^+ + q^- \quad (4)$$

$$2H^+ + 2q^- \rightarrow H_2(g) \quad (5)$$

### 3.5 Pyrocatalytic RhB degradation and mechanism

The pyrocatalytic performance of the KN-np catalyst was further characterized by its ability to degrade RhB organic dye under the same temperature fluctuation range (20−50 °C) following the same temperature curve represented in Fig. 5(a). RhB pyrocatalytic degradation was carried out for 16 successive thermal cycles, and the absorption spectra were recorded after every 2 thermal cycles. As shown in Fig. 7(a), the maximum intensity of the RhB absorbance at 554 nm decreases gradually with the thermal cycling process. The relative concentration $C/C_0$ of the dye solution derived from the experiment was plotted as a function of thermal cycles and represented in Fig. 7(b). In this plot, C and $C_0$ are the residual and the initial concentrations of the RhB solution. The degradation efficiency of RhB using KN catalyst reached around 84 % after 16 thermal cycles. The presence of KN catalyst is necessary for the degradation of RhB, as evidenced by the blank test, which demonstrates that after 16 thermal cycles, RhB dye degradation can be neglected, proving that temperature fluctuations cannot decompose RhB without the catalyst. These results confirm the role of the pyroelectric effect in the RhB degradation process. Additionally, by fitting the experimental data points in Fig. 7(b) to a pseudo-first-order kinetic equation ($-\ln(C/C_0) = k \cdot t$), Fig. 7(c) shows that the rate of dye decomposition increases almost linearly with thermal cycling, indicating the effectiveness of KN-np towards RhB degradation.

The kinetic rate constant for RhB decomposition reached 0.11 per thermal cycle, the highest value compared with 0.008 $cycle^{-1}$ obtained previously with $BaTiO_3$ nanofibers using the same dye [74]. The $BaTiO_3$@2.5%ZnO heterostructure showed a constant RhB dye degradation rate of 0.064 per thermal cycle when subjected to 42 thermal cycles ranging from 30 to 54 °C [21]. Regarding alkaline niobate, $NaNbO_3$ nanosheets demonstrated a degradation efficiency of 76 % after 24 thermal cycles under 23−50 °C heating and cooling cycles [67]. The high polarization of KN-based materials is responsible for their strong pyrocatalytic, piezocatalytic, and photocatalytic responses, as reported in several studies [33,34,75,76]. Otherwise, poling treatment has been shown to significantly enhance the pyrocatalytic decomposition of RhB dye using $K_{0.5}Na_{0.5}NbO_3$ bulk material undergoing 100 cold-hot cycles between 30 and 60 °C [75]. However, a poling process is unnecessary when using nano/micro pyroelectric powders since each particle is considered a single-oriented polarization domain. Zhang et al. [56] demonstrated that the strong local polarization of orthorhombic $KNbO_3$ nanowires contributes to superior photocatalytic activity for RhB decomposition compared to their monoclinic counterparts. To our knowledge, this is the first study to explore orthorhombic KN-np for pyrocatalytic RhB dye degradation and HER. The promising results provide a solid foundation for further investigation of KN ferroelectric material with various morphologies and crystalline structures in future pyrocatalytic applications. For clarity, Table 2 compares the pyrocatalytic performance of KN-np in the degradation of RhB with that of other pyroelectric materials.

Table 2. Comparison of the pyrocatalytic performance of different pyroelectric materials for RhB (5 mg $L^{-1}$) dye degradation.

| Pyro-catalysts | Cycles | Cycling temperature (°C) | Degradation efficiency (%) | Kinetic rate constants per cycle (× $10^{-3}$) | Refs. |
|---|---|---|---|---|---|
| KN-np | 16 | 20–50 | 84 | 110 | This work |
| $K_{0.5}Na_{0.5}NbO_3$ | 100 | 30–60 | 3.3 | 0.41 | [75] |
| $NaNbO_3$ nanosheets | 24 | 23−50 | 76 | − | [67] |
| $NaNbO_3$ nanocubes | 24 | 23−50 | 33 | − | [67] |
| $NaNbO_3$ nanofibers | 24 | 15–50 | 63 | − | [20] |
| $BaTiO_3$ nanofibers | 72 | 30–52 | 48 | 7.69 | [74] |
| $BaTiO_3$ crystals | 140 | 28–60 | 8 | 0.61 | [77] |
| $BaTiO_3$@2.5%ZnO | 42 | 30–54 | 97 | 60 | [78] |

Assessing the catalyst's reusability is crucial to determining its potential practical applications. In this respect, the reusability of the KN-np catalyst was effectively demonstrated by the repeated degradation of RhB with the recycled catalyst. As Fig. 7(d) shows, a comparable activity level was maintained throughout the first three cycles, with a slight decrease in the fourth cycle, proving good catalyst stability.

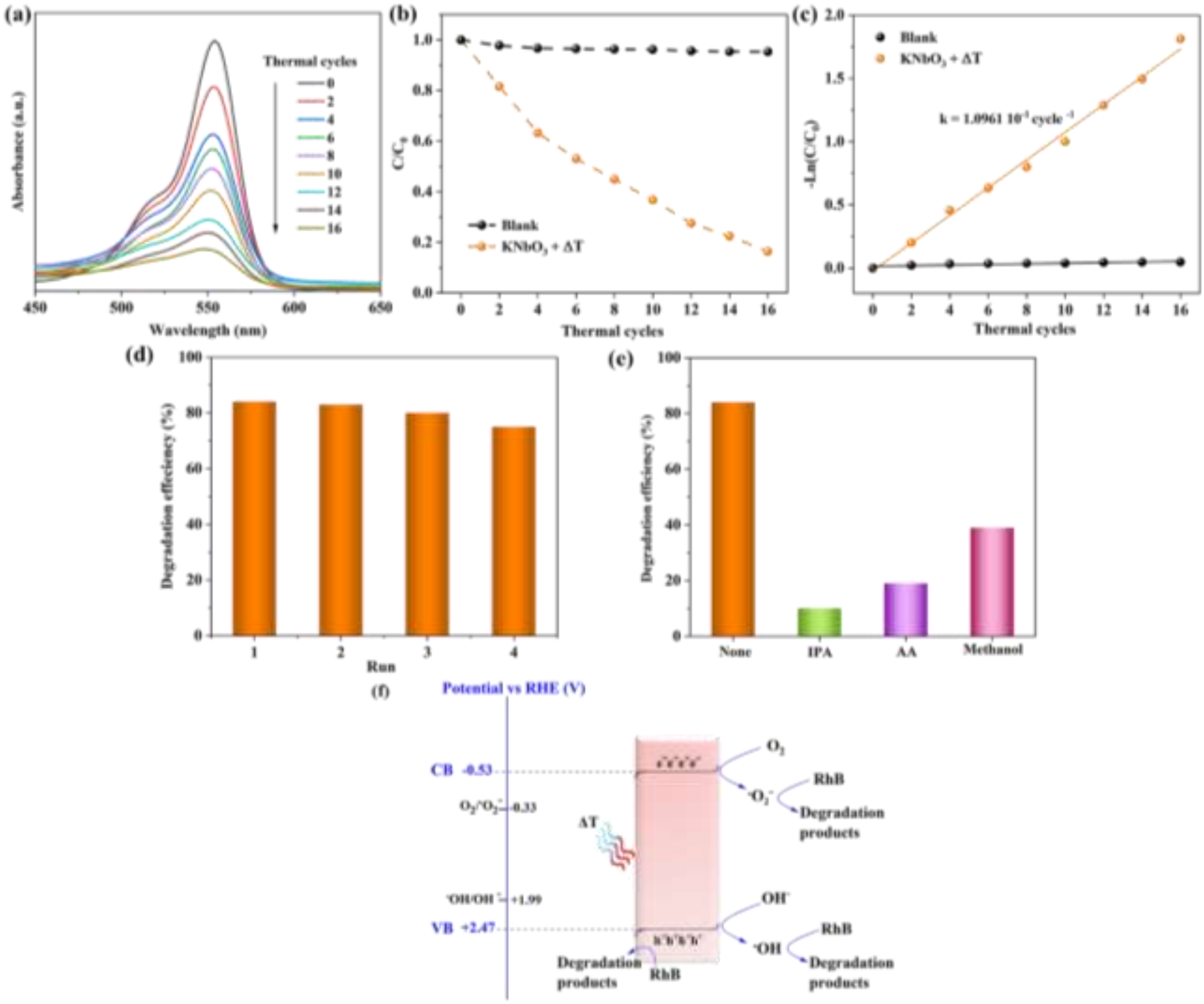

Fig. 7. (a) Absorption spectra of RhB dye solution separated from KN-np catalyst suspension after experiencing thermal cycling from 20 to 50 °C; (b) Variation of the relative RhB concentration vs. thermal cycles; (c) Plots of -ln($C/C_0$) vs. thermal cycles for the studied pyro-catalyst; (d) Cycling runs for RhB pyro-degradation; (e) Pyrocatalytic degradation of RhB by KN-np under thermal cycling in the presence of scavengers; (f) Schematic illustration of the possible pyrocatalytic mechanism of RhB degradation using KN-np catalyst.

To elucidate the pyrocatalytic mechanism underlying the decomposition of the RhB dye using the KN-np catalyst in cold-hot cycles, charge-trapping experiments were carried out by introducing different scavenging agents. This follows the same process as pyrocatalytic evaluation, apart from involving radical scavengers. The test involves the detection of hydroxyl radicals ($^{\bullet}OH$), superoxide radical anions ($^{\bullet}O_2^-$), and pyro-induced holes ($h^+$) [3]. As shown in Fig. 7(e), the degradation efficiency of RhB was significantly decreased and fell to a minimum when IPA was added over 16 cold-hot cycles.

Similarly, adding AA as a superoxide radical anion ($^{\bullet}O_2^-$) quencher also showed observable inhibition of the pyrocatalytic degradation of RhB. Furthermore, the least amount of the dye's pyrocatalytic decomposition was lost when methanol was added as an $h^+$ scavenger. In a nutshell, the charge trapping experiments demonstrate that the pyrocatalytic decomposition of RhB using the KN-np catalyst was triggered by pyro-induced hydroxyl radicals ($^{\bullet}OH$), superoxide ($^{\bullet}O_2^-$), and radicals, holes ($h^+$) [75].

From above, a mechanism of the pyroelectric catalytic degradation is proposed. When KN pyroelectric nanoparticles are exposed to temperature fluctuations, a redistribution of charge carriers at the particle surface occurs due to the temperature dependence of polarization charges [57]. The introduced polarization field drives free electrons and holes toward the opposite crystalline surfaces, where they are attracted and accumulated. At these surfaces, the electrons and holes interact with dissolved oxygen and hydroxyl ions, generating reactive species such as $^{\bullet}O_2^-$ and $^{\bullet}OH$ radicals (Fig. 7(f)). The high pyroelectric potential in the KN-np catalyst prompts electrons and holes to participate more effectively in the redox reaction [75]. On the one hand, oxidative decomposition can occur through the direct interaction of holes with adsorbed organic molecules or their capture by $OH^-$ ions, leading to the formation of reactive $^{\bullet}OH$ radicals [7]. On the other hand, reductive decomposition is likely facilitated when dissolved oxygen captures free electrons, generating $^{\bullet}O_2^-$ radicals [7]. Afterward, when the electric field of the accumulated charges counterbalances the polarization field in the KN-np catalyst, a new equilibrium is formed, slowing down the oxidation-reduction reactions. Like the pyroelectric current generation mechanism, decreasing temperature increases the polarization, which disrupts this equilibrium and triggers reverse charge transfer and new redox reactions. The possible reactions involved in the pyrocatalytic degradation of the RhB dye are described below in equations (6-10). Pyroelectric catalysis mechanism works the same way in nanostructures, bulk materials and films. However, the efficiency of this process is often significantly lower in bulk materials and thin films, due to their smaller surface area exposed to the dye solution, which limits the interaction sites. Furthermore, the high dispersion capacity of the KN catalyst enhances its surface interactions with the RhB dye, making it the best choice for the pyrocatalytic decomposition of dyes.

$$KNbO_3 \xrightarrow{\Delta T} KNbO_3 + e^- + h^+ \tag{6}$$

$$O_2 + e^- \rightarrow {}^{\bullet}O_2^- \tag{7}$$

$$OH^- + h^+ \rightarrow {}^{\bullet}OH \tag{8}$$

$$H_2O + h^+ \rightarrow {}^{\bullet}OH + H^+ \tag{9}$$

$$RhB + (h^+, {}^{\bullet}OH, {}^{\bullet}O_2^-) \rightarrow \text{Degraded products} \tag{10}$$

## 4 Conclusions

Orthorhombic $KNbO_3$ perovskite nanoplatelets were hydrothermally synthesized and, for the first time, used in a pyrocatalytic process to produce hydrogen and degrade rhodamine B dye by utilizing thermal energy from temperature changes. Structural, morphological, and piezoelectric properties, as well as pyro-electrochemical characteristics, were studied. Under cyclic heating-cooling between 20 and 50 °C, $KNbO_3$ nanoplatelets, exhibiting significant polarization, generated a high hydrogen yield of 680 $\mu mol \cdot g^{-1}$ after 30 thermal cycles, with a large hydrogen generation rate of approximately 22.67 $\mu mol \cdot g^{-1}$ per thermal cycle. The negative potential of the conduction band minimum of $KNbO_3$ nanoplatelets relative to the $H^+/H_2$ level enables efficient hydrogen evolution. Moreover, $KNbO_3$ nanoplatelets with good pyrocatalytic activity enable efficient RhB dye degradation in cold-hot cycles ranging from 20−50 °C, highlighting their potential for environmental remediation. The degradation efficiency of RhB reached 84 % after only 16 thermal cycles with a high kinetic rate constant of 0.11 per thermal cycle. The possible pyrocatalytic mechanisms have been proposed based on the results of each of both experiments. This environmentally friendly and efficient pyrocatalytic process demonstrates a promising pathway for sustainable hydrogen production and pollutant degradation driven by thermal energy using the lead-free $KNbO_3$ catalyst. Our findings highlight the strong polarization and good pyroelectric properties of $KNbO_3$ nanoplatelets, which enhance their catalytic activity, providing a viable route for pyrocatalysis

**Acknowledgments**

This work was supported by the Horizon Europe Framework Program Action HORIZON-MSCA-2022-SE-01-H-GREEN (No. 101130520), the Region of Hauts-De-France (HDF), and the Slovenian Research Agency (research core funding P2-0105, P1-0125, project J2-60035 and Young Researcher project). Jena Cilenšek and Val Fišinger are gratefully acknowledged for their help in the laboratory.